\documentclass[aps,pra,twocolumn,superscriptaddress,showpacs,amsmath,amssymb]{revtex4}

\usepackage[dvips]{graphicx}
\usepackage[latin2]{inputenc}

\usepackage{amsfonts}

\begin{document}

\title{Reversible and irreversible dynamics of a qubit interacting with a small environment}

\author{Giuliano Benenti}
\affiliation{CNISM, CNR-INFM \& Center for Nonlinear and Complex systems,
Universit\`{a} degli Studi dell'Insubria, via Valleggio 11,
I-22100 Como, Italy \\ \& Istituto Nazionale di Fisica Nucleare, Sezione di Milano, via Celoria 16, I-20133 Milano, Italy}

\author{G. Massimo Palma}
\affiliation{ NEST - CNR (INFM) \& Dipartimento  di Scienze Fisiche ed Astronomiche, Universit\`a di Palermo, via Archirafi 36,
I-90123 Palermo, Italy}

\date{\today}


\begin{abstract}
We analyze the dynamics of a system qubit interacting by means a sequence of pairwise collisions with an environment consisting of
just two qubits.  We show that the density operator of the qubits approaches a common time averaged equilibrium state,
characterized by large fluctuations, only for a random sequence of collisions. For a regular sequence of collisions the qubit
states of the system and of the reservoir undergo instantaneous periodic oscillations and do not relax to a common state.
Furthermore we show that pure bipartite entanglement is developed only when at least two qubits are initially in the same pure
state while otherwise also genuine multipartite entanglement builds up.
\end{abstract}

\maketitle
\section{Introduction}
Quantum information is providing new perspectives and is opening new issues in the analysis of the dynamics of open
quantum systems. For instance in \cite{PalmaSE96} the loss of coherence of a system of qubits has been shown to be linked to the
irreversible flow of information between the system and the bath due to a build up of entanglement between the two. Such model has
been extended to the case in which couplings between the bath degrees of freedom are present \cite{Tessieri, Paganelli, Mancini,
Milburn}. In particular it was shown that, due to the monogamy of entanglement \cite{CKW}, the presence of intrabath entanglement
can inhibit decoherence processes. The irreversible dissipation of a qubit has been analyzed in \cite{Scarani, Ziman, Ziman2} in
terms of an exchange of information by means of repeated collisions between a single qubit and a reservoir of  an arbitrarily
large number $N$ of identical independent qubits all prepared in the same state.
 In most of the existing literature an irreversible dynamics is obtained under the assumption that the reservoir has a large number of
 degrees of freedom. For instance,
in the derivation of master equations the Born-Markov approximation is justified on the basis of a weak coupling with a basically
unaffected large reservoir. On the other hand the dynamics of a system interacting with a small environment can have rich new
features ranging from memory effects, Poincar\'e recurrences etc. . Our work is somehow in the same spirit of the works by Mahler
and coworkers \cite{mahler}, in which the interaction of various quantum systems with finite environments, like finite quantum
networks or quantum Turing machines, has been analyzed with particular focus on the onset of quantum chaos. In the present paper
we will address the problem whether, and under which conditions, an irreversible dynamics can appear when a system interacts with
very small reservoirs, a situation often encountered in the physics of mesoscopic systems \cite{small1} and of quantum information
processing \cite{small2,small3} and of growing interest. To this goal we will use a repeated collision model to analyze the time
evolution of a qubit interacting with the smallest non trivial reservoir consisting of just two qubits. In spite of the apparent
simplicity of the model, the system dynamics shows interesting new features. Due to the small size of the environment, the reduced
dynamics of the system is characterized by large fluctuations. However an irreversible dynamics is retrieved when the system
dynamics is averaged over a sufficiently large number of collisions (as we will discuss later this is different from the usual
time coarse grained introduced in the derivation of a master equation). We will show that a - time averaged - equilibrium state is
reached only for a random sequence order of collisions. This is not a-priori an obvious result. It is important to stress that the
reduced dynamics of the system qubit cannot, in this case, be described in terms of a Markovian master equation, as we will show
below.

\section{The model}
To set the scenario and to illustrate the approach of our work let us review the repeated collision model. Consider a set of $N+1$
qubits, the first of which is the system qubit and the remaining $N$ are the reservoir. The interaction between system and
environment is due to pairwise  collisions between the system and a singe reservoir qubit. Each collision is described in terms of
a unitary  operator $U_{i}$. After $t$ collisions the overall state of the system plus reservoir  is
\begin{equation}
\varrho_{S E}^{(t)} = U_{i_t} \cdots U_{i_2} U_{i_1} \varrho_{S E}^{(0)}  U_{i_1}^{\dagger}U_{i_2}^{\dagger} \cdots
U_{i_t}^{\dagger},
\label{Un}
\end{equation}
where $\varrho_{S E}$ is the total density operator and the sequence of labels $i_1 \cdots i_t$ specify the order with which the
environment qubits collide with the system one. In general each collision modifies the entanglement between the system and the
environment qubits, in particular initially separable qubits will become entangled. Such model was considered to analyze the
processes of thermalization \cite{Scarani} and of homogenization \cite{Ziman,Ziman2}. These were shown to be achieved when the
collisions are described by the partial swap operator
\begin{equation}
U_{i} = \cos \eta \openone + i\sin\eta S_i,
\end{equation}
where $S_i$ is the swap operator between the system qubit and the $i$th environment qubit defined by the relation $S_i
|\psi\rangle \otimes |\phi\rangle = |\phi\rangle \otimes |\psi\rangle$ (here $|\psi\rangle \otimes |\phi\rangle$ is an arbitrary
product of single-qubit pure states). The environment was assumed to consist of a set of $N$ qubits all initially prepared in the
(generally mixed) state $\xi$. For instance in \cite{Scarani} $\xi$ is a thermal state of a single spin. The initial density
operator was therefore $\varrho_{SE}^{(0)} = \varrho_{S}^{(0)}\otimes \xi^{\otimes N}$. The number $N$ of environment qubits was
assumed to be infinite in order to neglect repeated collisions between the system qubit and the same environment qubit. Under
these assumption the system qubit was shown to relax to thermal equilibrium. More in general such model describes a homogenization
process in which the qubit system approaches state $\xi$ \cite{Ziman}. Such relaxation mechanism can be understood easily if one
notes that this models contains all the ingredients used in standard derivations of a master equation: the system interacts with
an environment which rapidly "forgets" the effects of the coupling with the system and returns to its stationary state.

Here we will look at the model from an entirely different viewpoint analyzing the system dynamics in the opposite limit of the
smallest non trivial environment, namely the limit $N=2$. The reason of this choice is twofold. On the one hand, although
thermalization or homogenization are clearly not possible for such small environment we would like to know if, and under which
conditions, the system and the environment reach some form of equilibrium. Furthermore, since the entanglement dynamics among the
three qubits can be followed and characterized in detail, it is possible to analyze under which conditions bipartite entanglement
and genuine tripartite entanglement builds up.

\section{Small environment}
The major new feature of the small reservoir limit is clearly the fact that the system qubit collides repeatedly with the same
environment qubits. In our case Eq.~(\ref{Un}) reduces to
\begin{equation}
\varrho_{012}^{(t)} = U_{i_t} \cdots U_{i_2}U_{i_1} \varrho_{012}^{(0)}
U_{i_1}^{\dagger}U_{i_2}^{\dagger} \cdots U_{i_t}^{\dagger},
\end{equation}
where the label $0$ refers to the system qubit while $1,2$ refer to the environment qubits.  We will not restrict ourselves to the
case in which $\varrho_{012}^{(0)}=\varrho_0 \otimes \varrho_{1}\otimes\varrho_{2}$ but we will consider also the case in which
some entanglement is initially present between the environment qubits. The order with which the $t$ collisions take place is
specified by the string of indices $i_1\cdots i_t$.  We have considered two limiting cases: a completely random sequence,
corresponding to the situation in which the system qubit collides with equal probability with each of the environment qubits, and
the regular periodic sequence $1212\cdots 12\cdots$, corresponding to the situation in which the system qubits collides
alternatively with the two environment qubits.

\subsection{Approach to equilibrium}
\begin{figure}
\begin{center}
\includegraphics[height=3.cm]{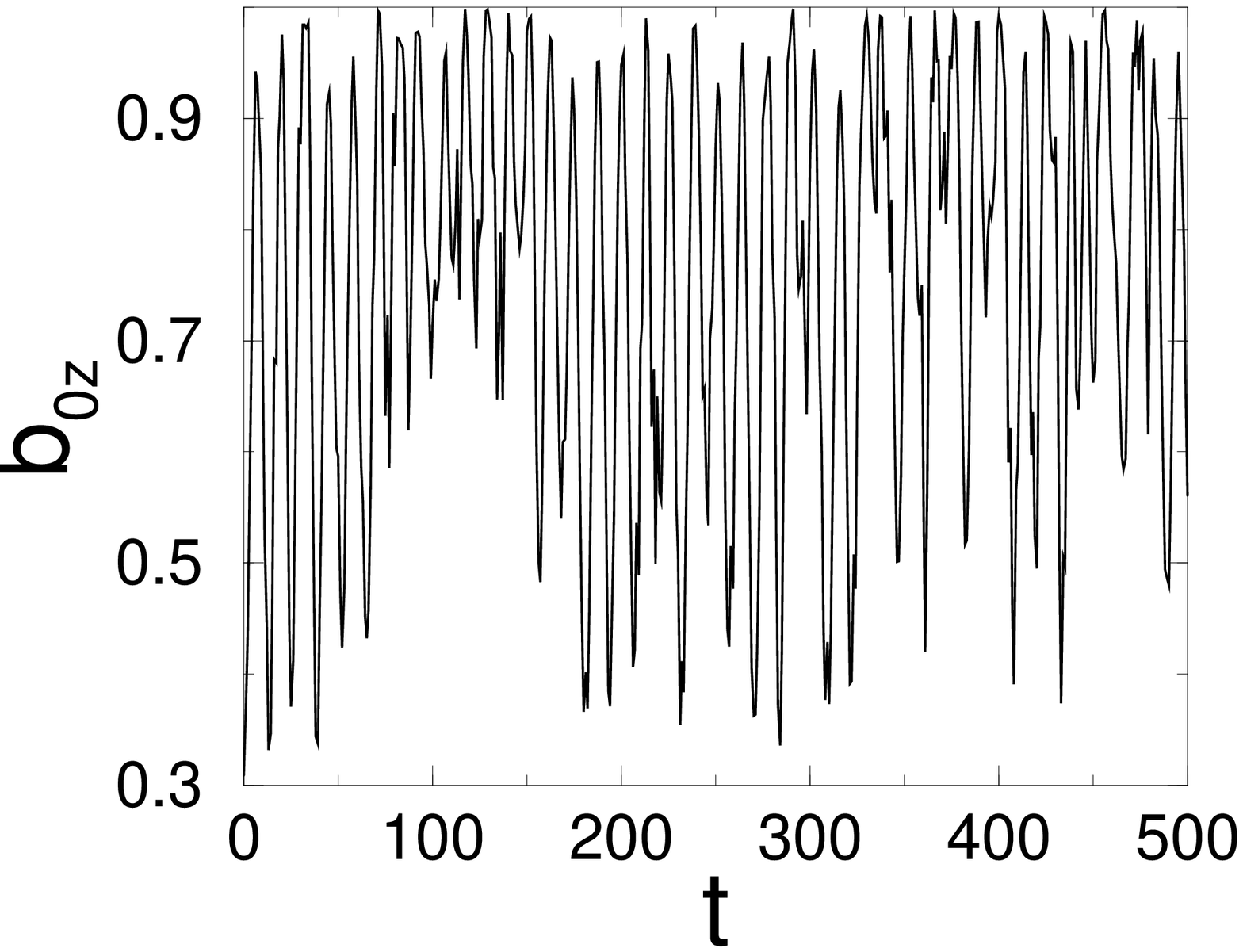}\hspace{0.0cm}
\includegraphics[height=3.cm]{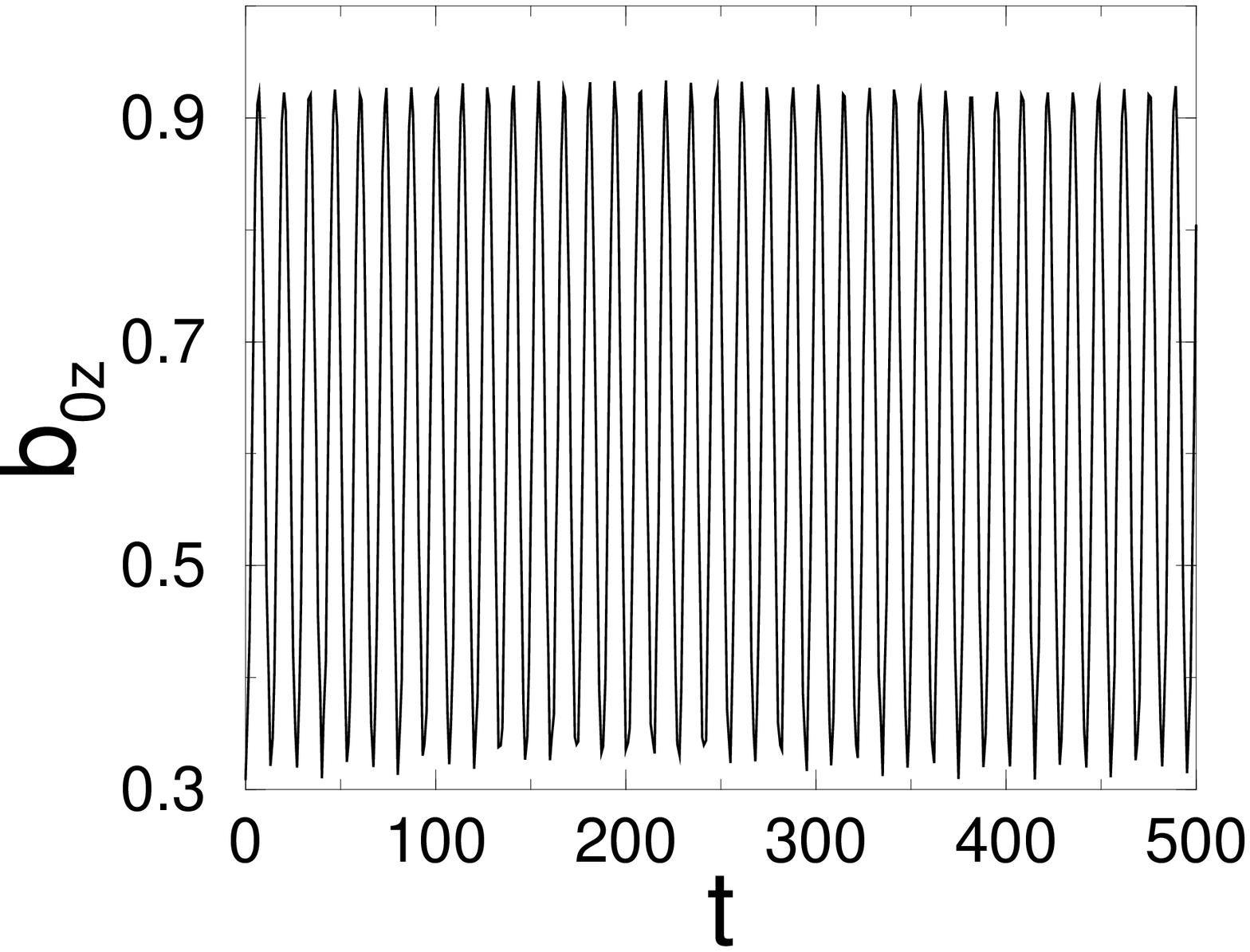}
\vspace{0.0cm}
\newline
\includegraphics[height=3.cm]{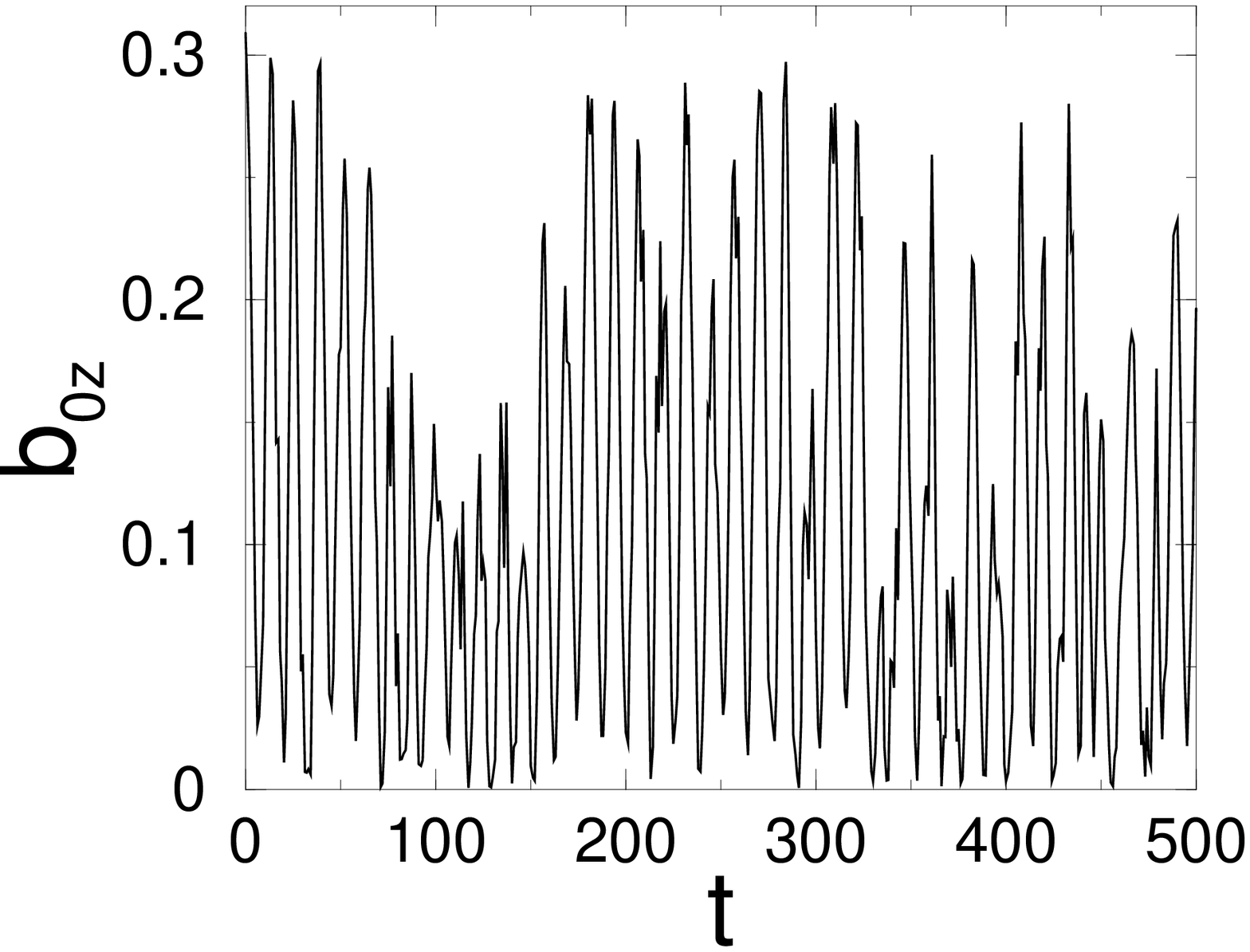}\hspace{0.0cm}
\includegraphics[height=3.cm]{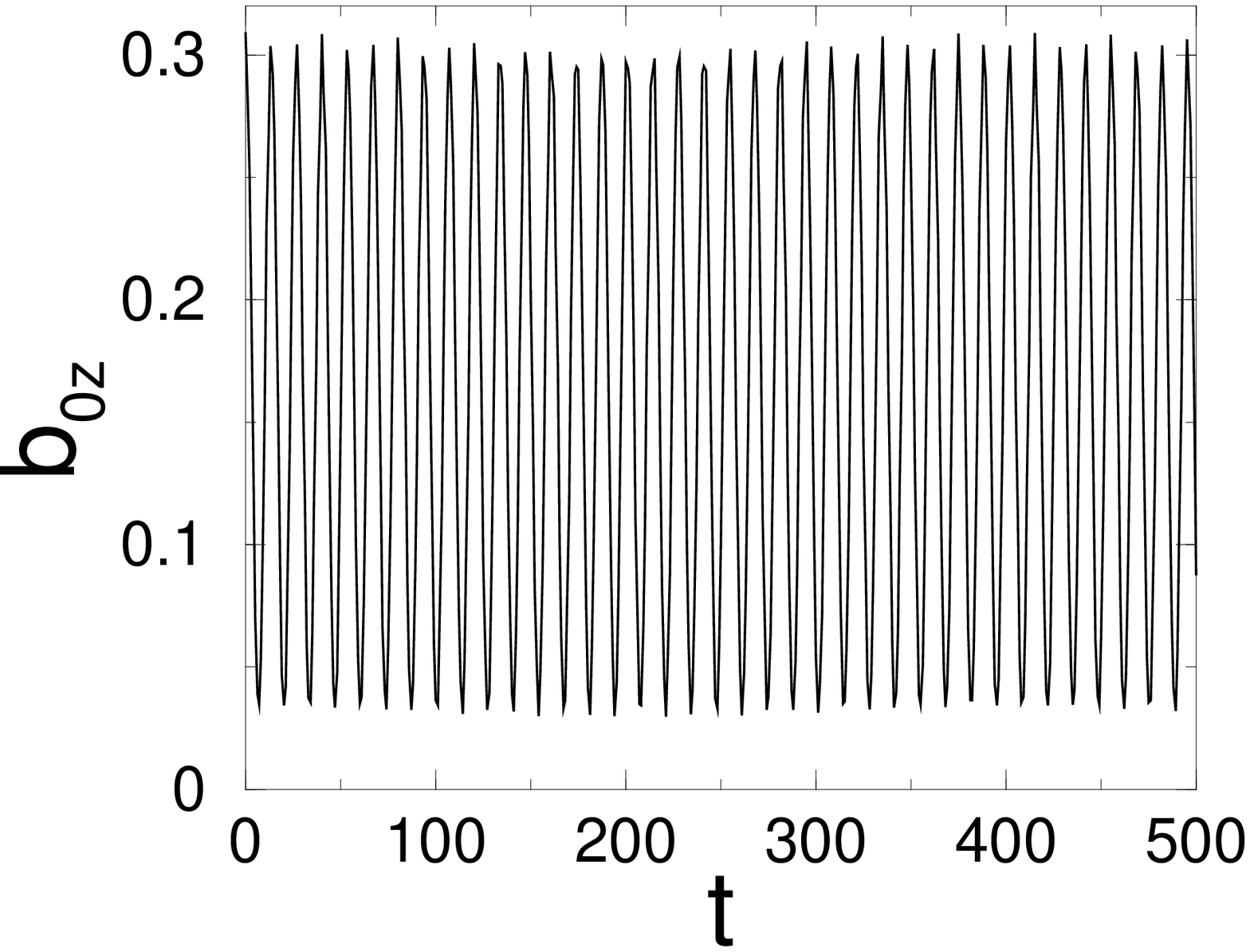}
\end{center}
\caption{Time evolution of the $z$-component of the
system qubit Bloch vector $\vec{b}_0$, for a random
(left) or an ordered (right) sequence of collisions.
The initial condition for the environment is the separable
state $|00\rangle$ (above) or the Bell state
$\frac{1}{\sqrt{2}}(|00\rangle+|11\rangle)$ (below).
Parameter values: $\eta=\frac{1}{10}\pi$,
$\theta=\frac{2}{5}\pi$.}
\label{fig1}
\end{figure}

In order to characterize the approach to equilibrium we will make use of an important feature of our system:  the total Bloch
vector, defined as ${\cal\vec {B}}  = \sum_{i=0}^2\vec{b}_i$, where $\vec{b}_i$ is $i$th qubit Bloch vector, is a constant of
motion. If the state of both system and environment qubits approach the same equilibrium reduced density operator, after a
sufficiently long number of collisions we must have $\vec{ b}_i \rightarrow {\cal\vec {B}} /3$. In Fig.~\ref{fig1} we show the
time evolution of the $z$-component of the system qubit Bloch vector $\vec{b}_0$, for a random and an ordered sequence of
collisions. A similar behavior is shown by the $x$ and $y$ components (data not shown). The system qubit is assumed to be in the
state $\cos\left(\frac{\theta}{2}\right)|0\rangle+ \sin\left(\frac{\theta}{2}\right)|1\rangle$ while the  environment is assumed
to be initially in the state $|00\rangle$ or in the Bell state $\frac{1}{\sqrt{2}}(|00\rangle+|11\rangle)$. Such choice has been
made to illustrate the new features of the dynamics for two different instances of environment in a pure state with an equal
reduced density operator for the two environment qubits and with different bipartite initial entanglement.

The figures clearly show that while for the ordered sequence of collisions the $z$-component of the Bloch vector exhibits periodic
oscillations, strong and rapid fluctuations appear for a random sequence of collisions, due to the extremely small size of our
reservoir. This is very different from what happens in the large reservoir limit, where, for sufficiently weak collisions, i.e.
for $\eta$ small enough, the environment qubits are hardly modified by the interaction with the system, while, after a
sufficiently large number of collisions the density operator of the system qubit becomes monotonically arbitrarily close to $\xi$
\cite{Scarani,Ziman}.

\begin{figure}
\begin{center}
\includegraphics[height=3.1cm]{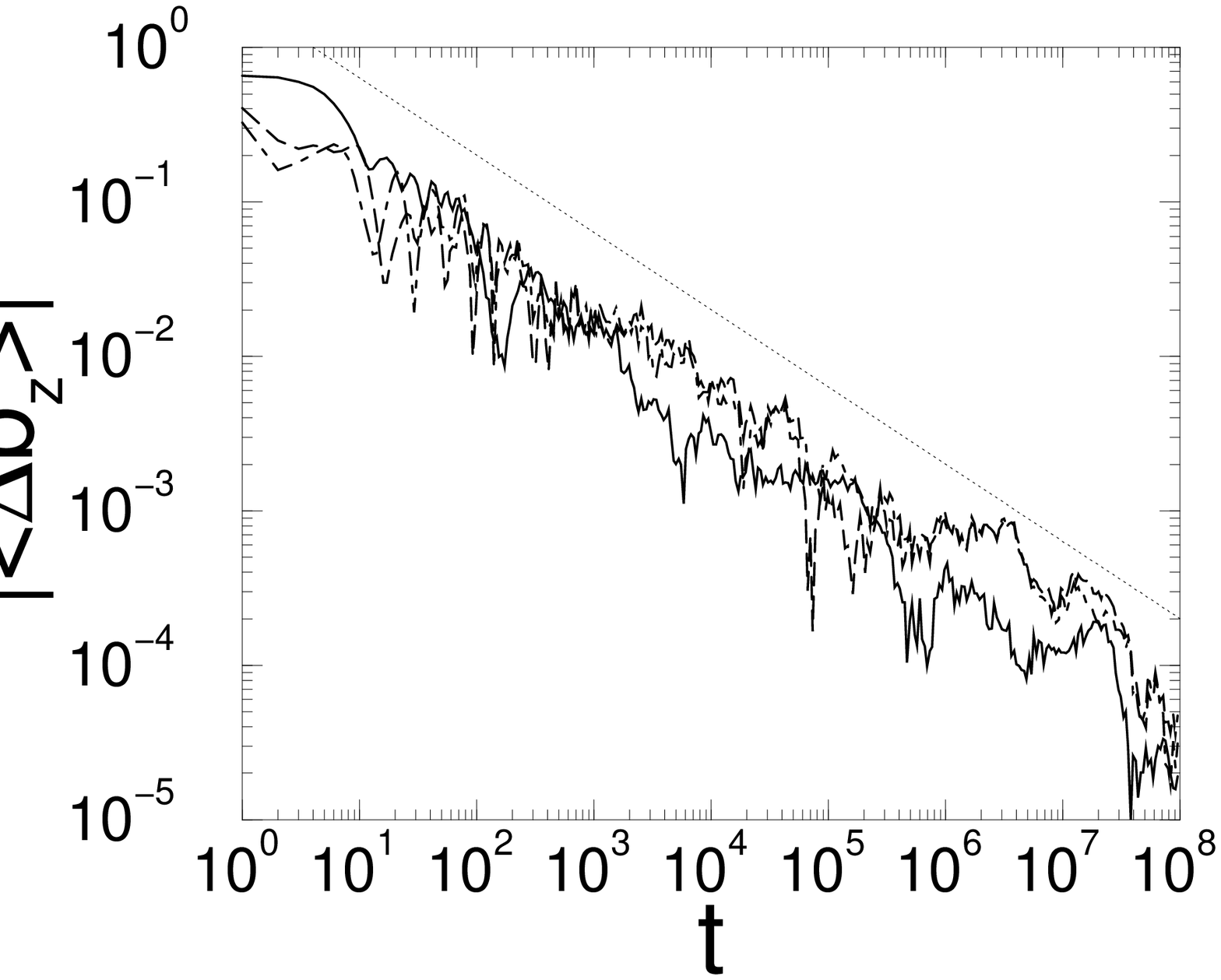}\hspace{0.0cm}
\includegraphics[height=3.1cm]{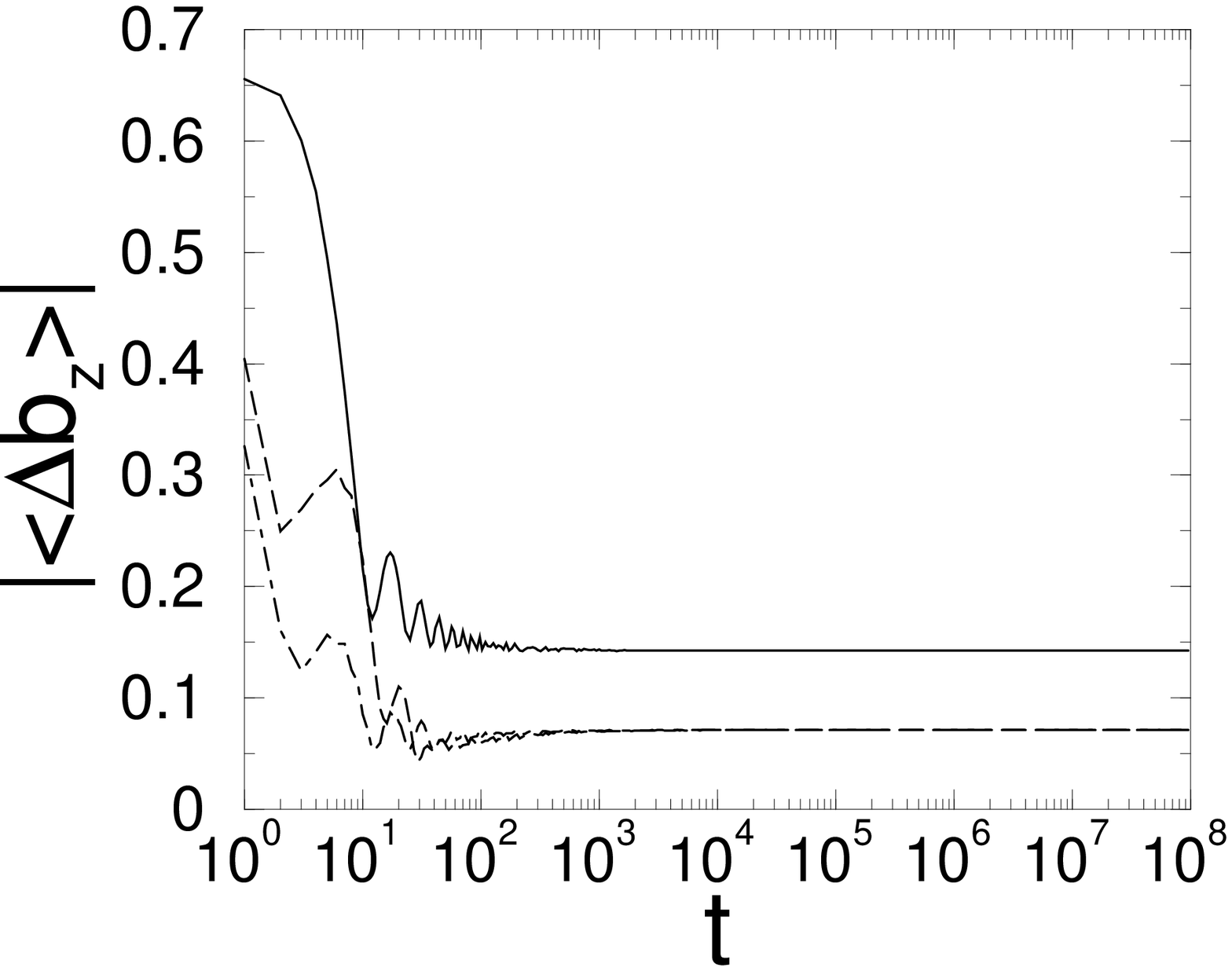}
\vspace{0.4cm}
\newline
\includegraphics[height=3.cm]{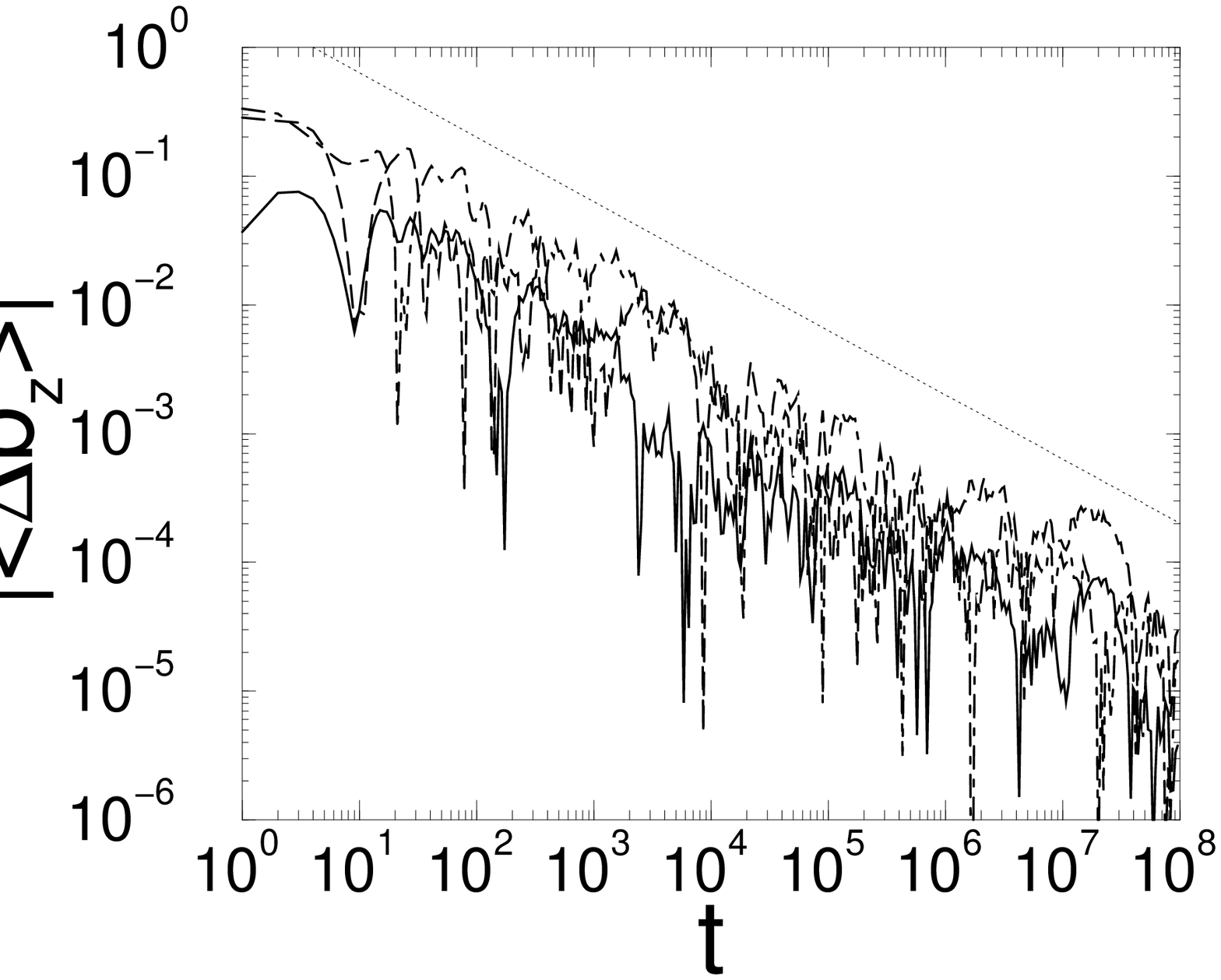}\hspace{0.0cm}
\includegraphics[height=3.cm]{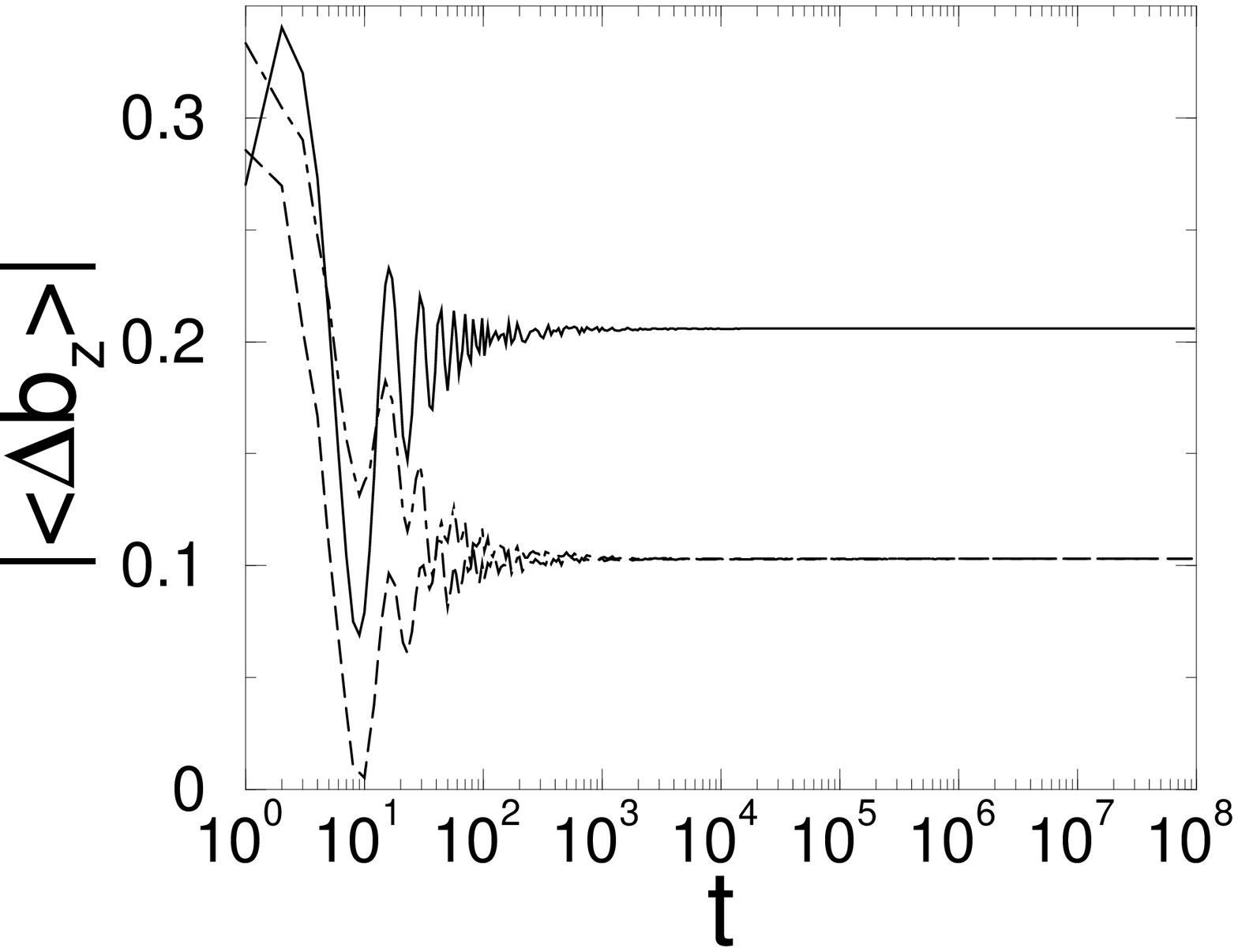}
\end{center}
\caption{Length $|\langle \Delta b_{iz} \rangle|$ of the $z$-component of the vector
$\langle \Delta \vec{b}_i\rangle$, for the system qubit ($i=0$,
solid curves) and the environment qubits ($i=1,2$, dashed and
dot-dashed curves, respectively). The straight lines show a decay
$\propto t^{-1/2}$. Same initial conditions and parameter values
as in Fig.~\ref{fig1}.}
\label{fig2}
\end{figure}

In our system a steady state is however reached if one looks at a time averaged dynamics. In order to explore the emergence of a
time-averaged equilibrium state, we have evaluated the components of the vector
\begin{equation}
\langle \Delta \vec{b}_i \rangle(t) = \frac{1}{t+1} \sum_{t^\prime=0}^t \left(\vec{b}_i(t^\prime)-\frac{\vec{\cal B}}{3}\right) =
\langle \vec{b}_i \rangle(t) - \frac{\vec{\cal B}}{3},
\end{equation}
where $\langle A \rangle (t)$ denotes the time average of the quantity $A$ from time $0$ to time $t$ and $i$ labels the qubits. In
Fig.~\ref{fig2} the  length $|\langle \Delta b_{iz}\rangle|$ of the $z$-component of the vector $\langle \Delta \vec{b}_i\rangle$,
for the system and the environment qubits is plotted, for the same initial conditions and parameter values as in Fig.~\ref{fig1}.
It is important to note that in the time-averaged dynamics one looses knowledge of the exact sequence of collisions. If the same
time average were done in the homogenization process described in \cite{Ziman}, \emph{all} sequence of collisions would give
origin to the same irreversible dynamics.  This is not at all the case in the small reservoir limit. Our results suggest that the
approach to a time average  equilibrium state depends on how random - i.e. how compressible in the Kolmogorov sense
\cite{Kolmogorov} - the string identifying the sequence of collisions is. When  a regular pattern exists in the collision sequence
then the time-averaged dynamics of the system and the reservoir qubits settle on different values. Note incidentally that while in
the large $N$ limit, thanks to the fact that the system qubit collides with "fresh" reservoir qubits, the system dynamics can be
analyzed in terms of completely positive (CP) maps forming a semigroup leading to an effective Lindblad time evolution
\cite{Ziman2}, this is not possible in our case, since, due to the repeated collisions among the same qubits, the state of
reservoir qubits involved in the collision changes each time and an irreversible dynamics is obtained only after time average. We
emphasize that our time average is different from the typical coarse graining introduced in the derivation of a
Markovian master equation. Such coarse graining amounts at looking at the system dynamics over a timescale longer than the time
needed to the reservoir to reset to its steady state. In the repeated collision model with an infinite reservoir model the
environment sets to its equilibrium value after each individual collision as the system collides with different fresh reservoir
qubits.

The straight lines in the left plots of Fig.~\ref{fig2} show a decay $\Delta b_{iz} \propto t^{-1/2}$ for a random sequence of
collisions. This implies that the cumulative sum obtained by adding the vectors $\Delta \vec{b}_i(t')= \vec{b}_i (t') -
\frac{\vec{\cal B}}{3}$ from time $t'=0$ to time $t'=t$ grows $\propto t^{1/2}$. Such diffusive growth is the same as for Brownian
motion or for coin-tossing sequences and suggests that, given $t_1$ and $t_2$, the vectors $\Delta \vec{b}_i(t_1)$ and $\Delta
\vec{b}_i(t_2)$ are in practice uncorrelated, provided $|t_1-t_2|$ is sufficiently large.

\begin{figure}
\begin{center}
\vspace{0.2cm}
\includegraphics[height=3.05cm]{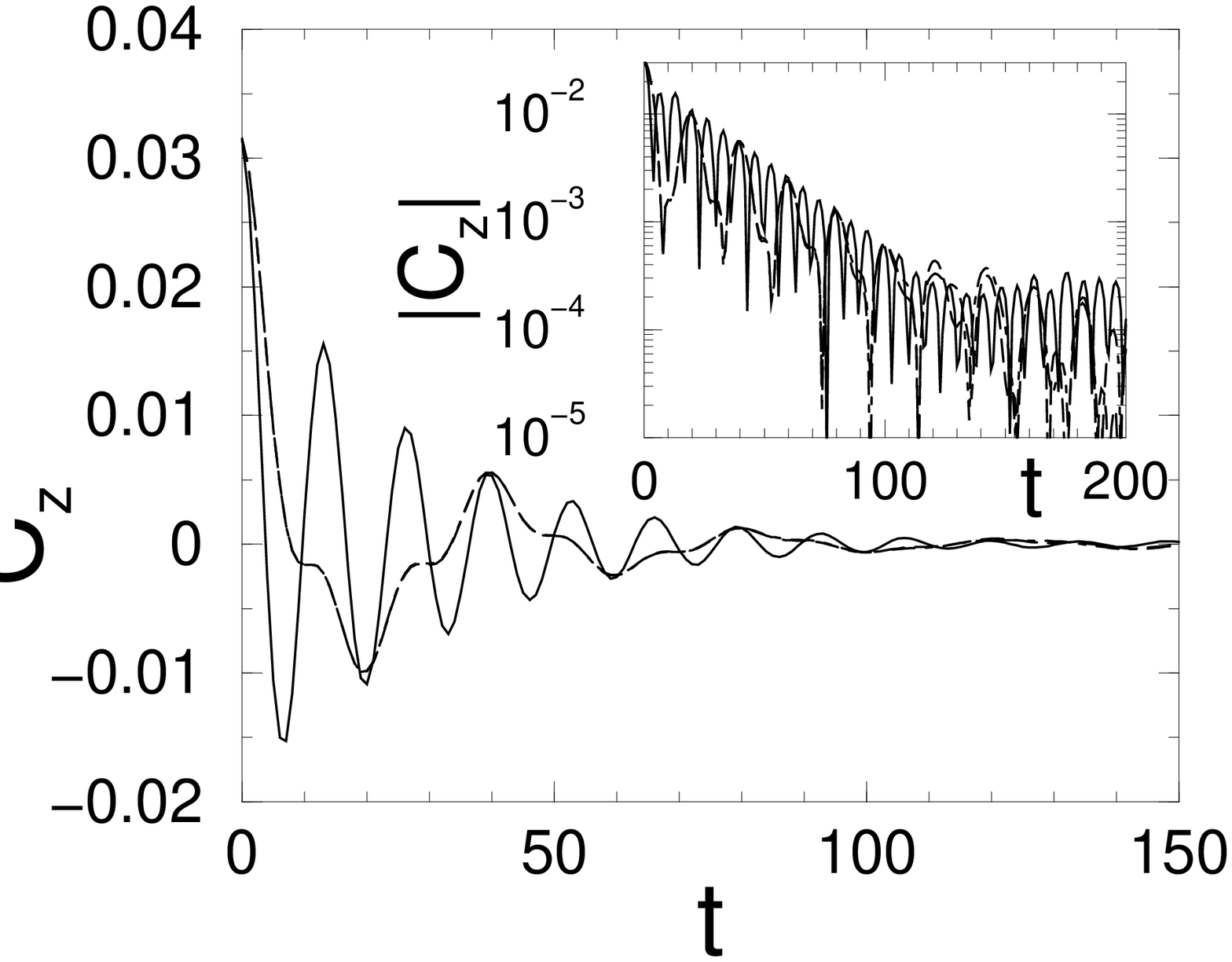}\hspace{0.0cm}
\includegraphics[height=3.05cm]{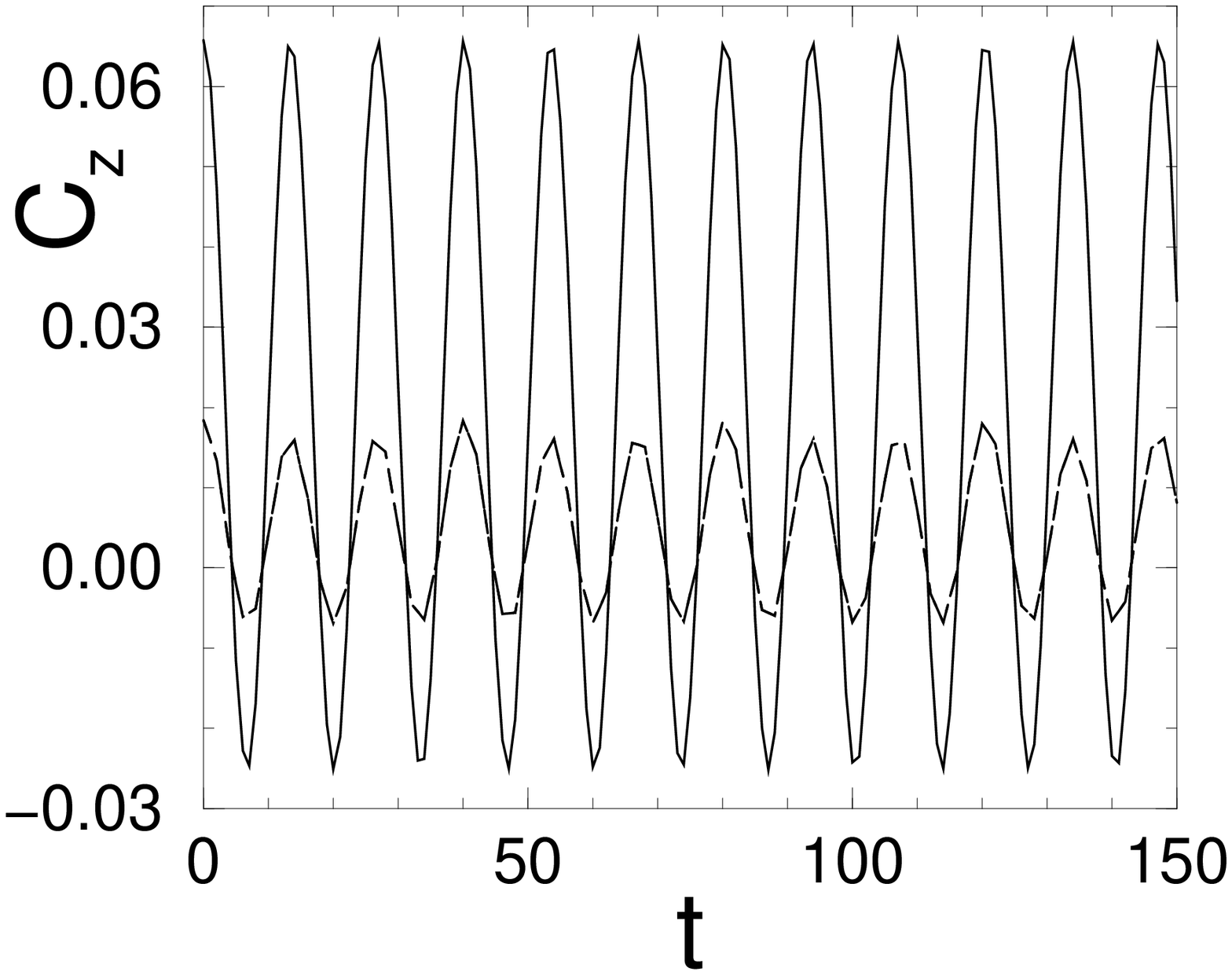}
\vspace{0.4cm}
\newline
\includegraphics[height=3.cm]{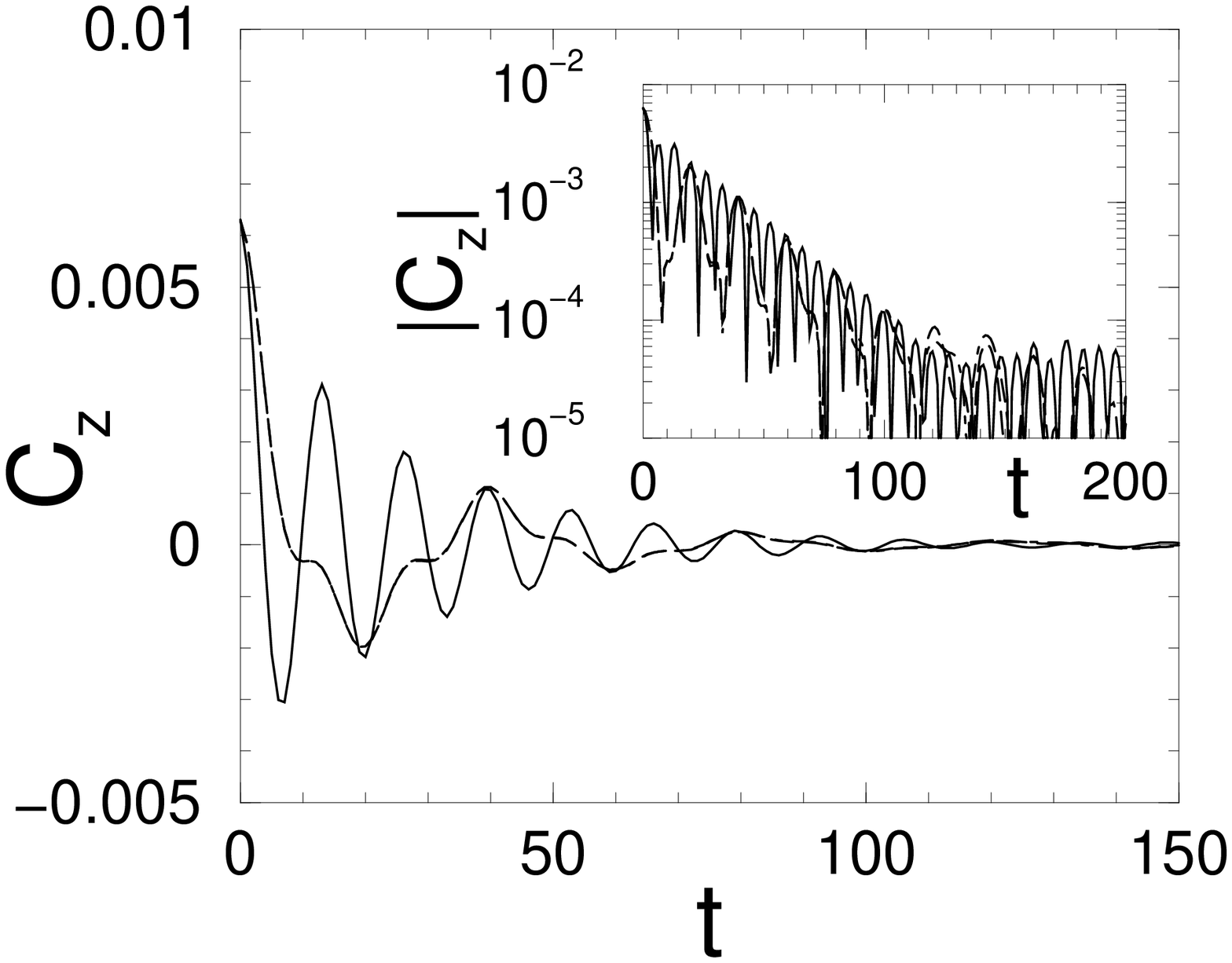}\hspace{0.3cm}
\includegraphics[height=3.cm]{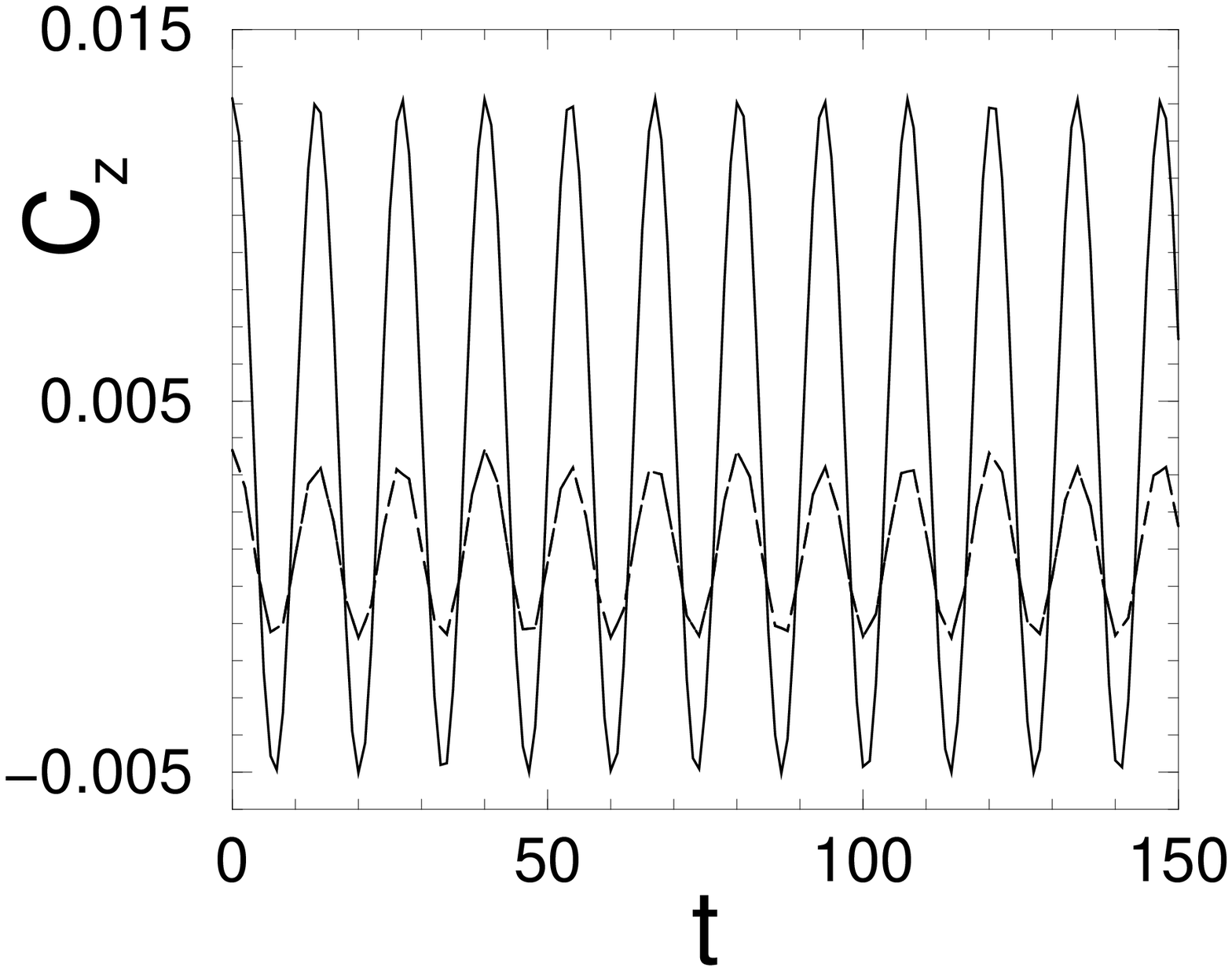}
\end{center}
\caption{Decay of the correlation function $C_{iz}$ (left plots,
corresponding to random sequence of collisions) versus oscillations
of the same correlation function in the case of a regular sequence
of collisions (right plots). Same meaning of the
curves as in Fig.~\ref{fig2}, initial conditions
and parameter values as in Fig.~\ref{fig1}.
Insets: $|C_{iz}|$ vs. time (semilogarithmic plot).}
\label{fig3}
\end{figure}

Further insight into the approach to equilibrium is gained by evaluating the self correlation defined as
\begin{equation}
\vec{C}_{ik}(t)=\lim_{T\to\infty}
\frac{1}{T+1}
\sum_{t^\prime=0}^T
\Delta b_{ik} (t^\prime) \Delta b_{ik} (t^\prime+t),
\end{equation}
where $k=x,y,z$.
In Fig.~\ref{fig3}, we show the decay of the correlation function $C_{iz}$ (left plots, corresponding to random sequence of
collisions) versus the oscillations of the same correlation function in the case of a regular sequence of collisions (right
plots). The insets show the exponential decay of $|C_{iz}|$ when the sequence of collisions is random. Note that $|C_{iz}|$
eventually oscillates around a value $\propto T^{-1/2}$ due to the finite sequence of data considered in computing the correlation
function (in these plots, $T=10^6$). This is again compatible with normal diffusive dynamics.


\subsection{Entanglement dynamics}
\begin{figure}
\begin{center}
\includegraphics[height=3.05cm]{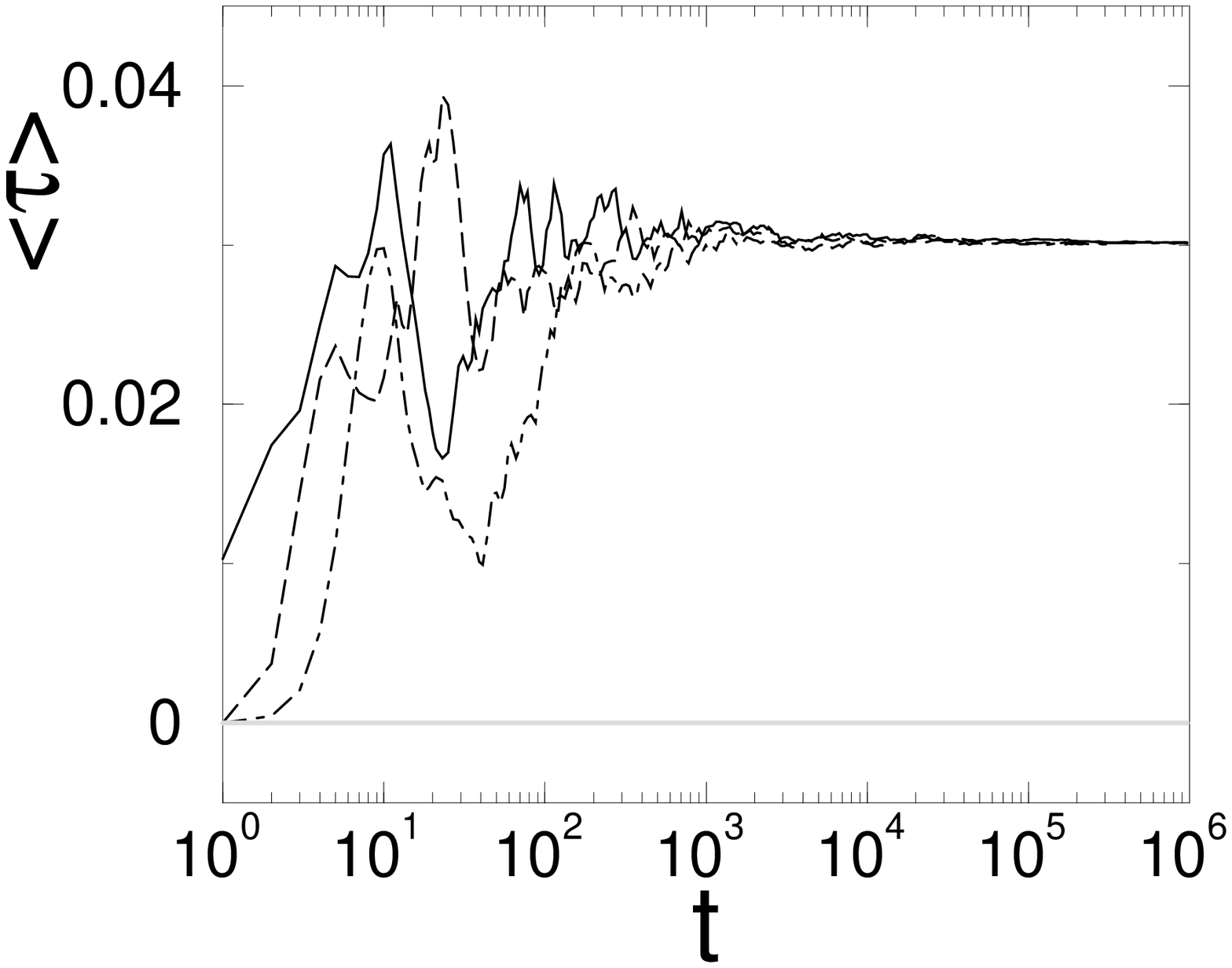}\hspace{0.0cm}
\includegraphics[height=3.05cm]{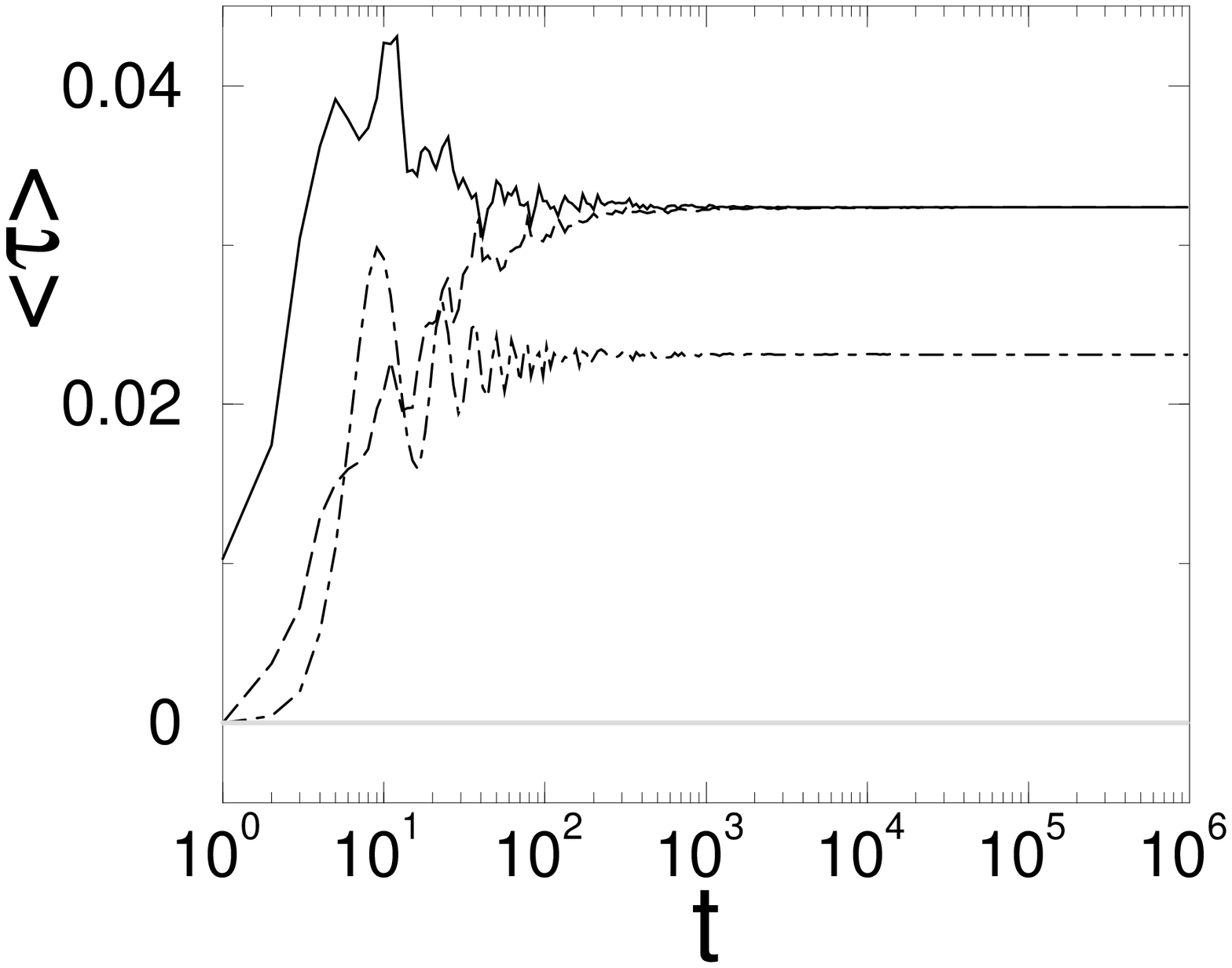}
\vspace{0.4cm}
\newline
\includegraphics[height=3.cm]{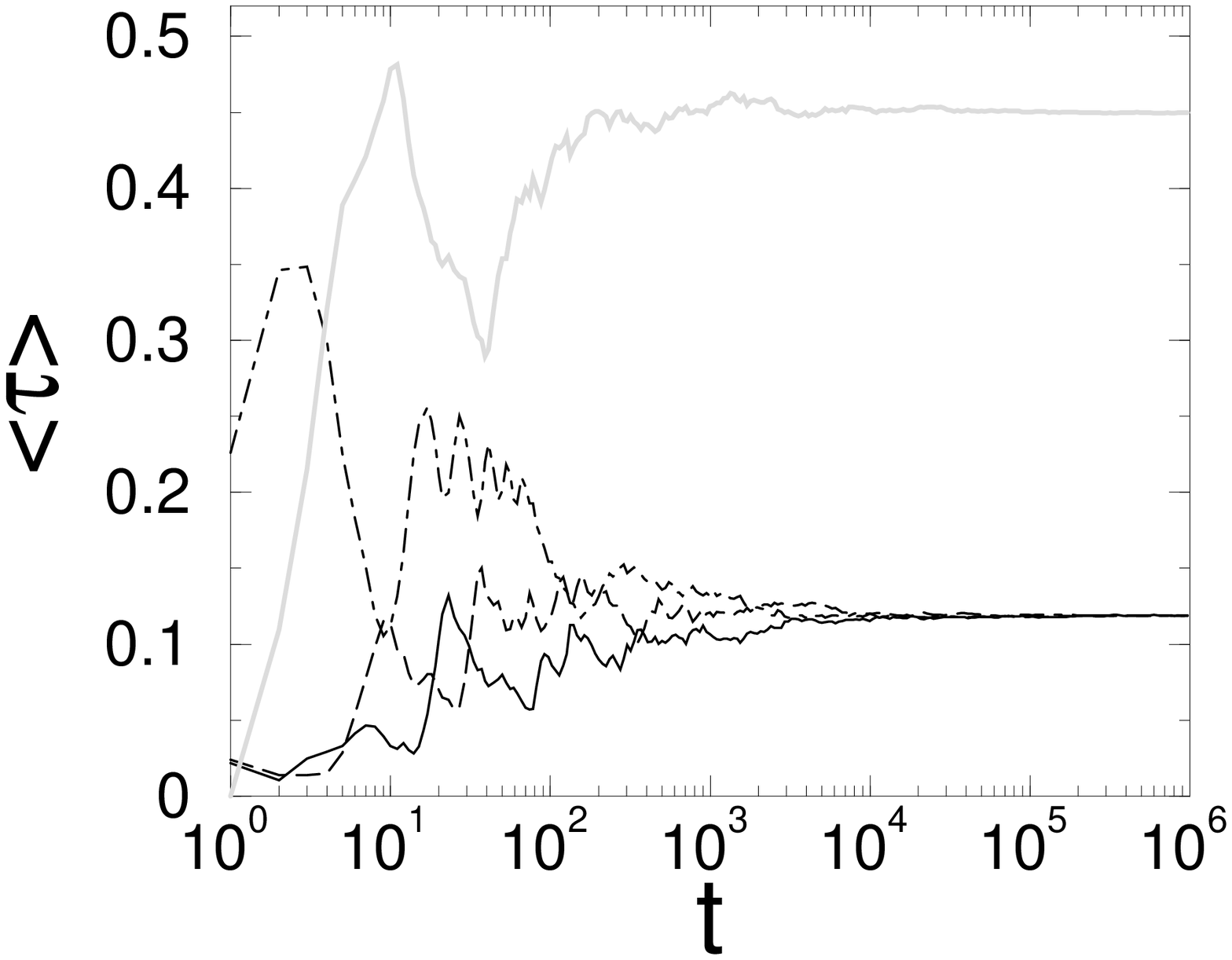}\hspace{0.0cm}
\includegraphics[height=3.cm]{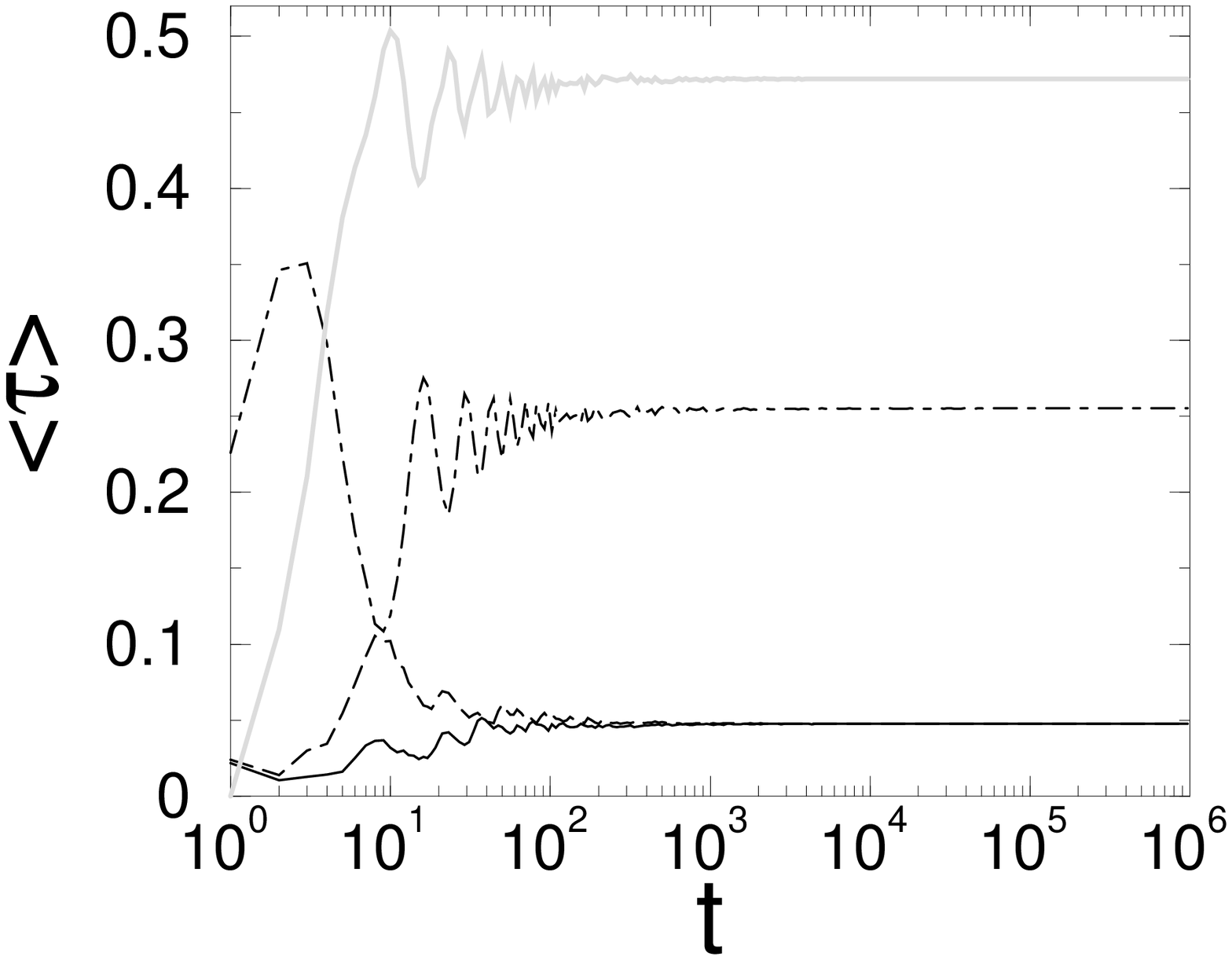}
\end{center}
\caption{Time-averaged tangles $\langle \tau_{0|1} \rangle$ (solid curves), $\langle \tau_{0|2} \rangle$ (dashed curves), $\langle
\tau_{1|2} \rangle$ (dot-dashed curves), and $\langle \tau_{0|1|2} \rangle$ (gray curves). Same initial conditions and parameter
values as in Fig.~\ref{fig1}.} \label{fig4}
\end{figure}

Interesting features of our model appear when the time evolution of bipartite and multipartite entanglement is analyzed. To this
end let us first introduce the tangle as a measure of bipartite entanglement. Given the density operator $\rho_{ab}$ of a
bipartite system of two qubits, the tangle $\tau_{a|b}$ is defined as
\begin{equation}
\tau_{a|b}(\rho)=[\max\left\{0,\alpha_{1}-\alpha_{2}-{\alpha_{3}}-{\alpha_{4}}\right\}]^2,
\end{equation}
where $\left\{\alpha_{i}\right\}$ ($i=1,..,4$) are the square roots of the eigenvalues (in non-increasing order) of the
non-Hermitian operator $\bar{\rho}=\rho(\sigma_{y}\otimes\sigma_{y})\rho^{*}(\sigma_{y}\otimes\sigma_{y})$, $\sigma_{y}$ is the
$y$-Pauli operator and $\rho^{*}$ is the complex conjugate of $\rho$, in the eigenbasis of $\sigma_z  \otimes\sigma_z$ operator.
In our model the tangle $\tau_{j|k}$ can be used to quantify the entanglement between the pair of qubits $i,j$ for an arbitrary
reduced density operator $\rho_{ij}$. Furthermore, when the overall state of the system is pure, the amount of entanglement
between qubit $j$ and all the remaining can be quantified by the tangle $\tau_{j| \mbox{rest}}=4\det\rho_j$. We have numerically
computed the tangles $\tau_{0|1}$, $\tau_{0|2}$, and $\tau_{1|2}$ of the two-qubit reduced density matrices and the three-tangle
$\tau_{i|j|k}=\tau_{i|jk}-\tau_{i|j} - \tau_{i|k}$, where $i,j,k$ can take values $0,1,2$ and where the tangle $\tau_{i|jk}$
measures the entanglement between the $i$th qubit and the rest of the system i.e. qubits $j,k$. The three-tangle $\tau_{0|1|2}$ is
a measure of the purely tripartite entanglement and is invariant under permutations of the three qubits \cite{CKW}. Our numerical
results of Fig.~\ref{fig4} show that, for a random sequence of collisions, the time-averaged tangles $\langle \tau_{0|1} \rangle$,
$\langle \tau_{0|2} \rangle$, and $\langle \tau_{1|2}\rangle$ saturate to the same limiting value.
This confirms once more the approach to statistical equilibrium for our system. Moreover, we can see from Fig.~\ref{fig4} that
genuine multipartite entanglement, i.e. $\langle \tau_{0|1|2} \rangle\ne 0$, is created when the two environment qubits are
initially prepared in a Bell state, but not if they are described by the same separable initial state. More generally, purely
bipartite entanglement builds up among the qubits when at lest two qubits (either both environment qubits or system and one
environment qubit) are in the same state and the initial state is a pure separable state. Indeed, if the initial state
$|\psi_{SE}^{(0)}\rangle$ is $|\psi \phi \phi\rangle$, or $|\phi \psi \phi\rangle$, or $|\phi \phi \psi\rangle$, then after $t$
collisions we obtain $|\psi_{SE}^{(t)}\rangle =\alpha(t) |\psi \phi \phi\rangle+ \beta(t) |\phi \psi \phi\rangle+\gamma(t) |\phi
\phi \psi\rangle$, where the coefficients $\alpha,\beta,\gamma$ are determined by the collision sequence. Using local unitary
transformations only i.e. with a tensor product of single qubit unitary transformations, we can map this state into
$a(t)|000\rangle+b(t)|100\rangle+c(t)|010\rangle+d(t)|001\rangle$, where $a,b,c,d\ge 0$ and $a^2+b^2+c^2+d^2=1$. This latter is
the standard form for the states of the $W$ class \cite{Cirac}, which are characterized by purely bipartite entanglement. However,
for general initial conditions, genuine multipartite entanglement builds up, as is the case when the two environment qubits are
initially in a Bell state. We point out however that the initial presence of bipartite entanglement in the reservoir is not a
necessary condition for the appearance of multipartite entanglement. Indeed, as should be clear from the symmetry argument
outlined above, a non zero time averaged three tangle $\langle \tau_{0|1|2} \rangle\ne 0$ builds up as long as the three qubits
are initially in different pure states. On the other hand if the initial state of the three qubits is of the {\it GHZ} family the
multipartite entanglement would remain purely tripartite. This of course is due to the fact that the partial swap operator does
not change the overall number of $|0\rangle$ and $|1\rangle$ states.

\section{Conclusions}

In summary, we have shown that relaxation (in time average) to statistical equilibrium is possible for a system of just three
qubits undergoing purely unitary evolution, provided that randomness is present in the sequence of pairwise collisions. We point
out that the results of this paper can be easily extended to the case in which the environment consists of a finite number $N$ of
qubits. In particular, a time-averaged equilibrium state characterized by purely bipartite entanglement is approached if the
initial state is separable and $N$ qubits are in the same state (either all $N$ environment qubits or the system and $N-1$
environment qubits). In this case, the $N+1$-qubit state evolves in a subspace 
of states of the $W$ class (spanned by the ``ground state'' $|0...0\rangle$
and the ``single-qubit excitations'' $|10...0\rangle,
|010...0\rangle,...,|0...01\rangle$), whose dimension
is $N+1$, much smaller than the overall dimension $2^{N+1}$ of the Hilbert space. Therefore, the equilibrium state can be attained
even though genuine multipartite entanglement is not developed. This implies that in a small, unitarily evolving quantum system statistical
relaxation is possible even though the dynamics is restricted to a subspace whose states are characterized by purely bipartite
entanglement. Since randomly chosen pure states in a many-qubit Hilbert space typically exhibit large multipartite entanglement
\cite{pascazio}, it follows that relaxation to equilibrium is possible without exploiting the full complexity of the Hilbert
space.

{\it Acknowledgments} G.B. acknowledges support from the PRIN 2005 " Quantum computation with trapped particle arrays, neutral and
charged". GMP acknowledges support from the PRIN 2006 "Quantum noise in mesoscopic systems".  We acknowledge useful discussion
with V.~Scarani and V.~Bu\v{z}ek.


\end{document}